\documentstyle[twoside,fleqn,espcrc2,epsf]{article}

\newcommand{\AmS}{{\protect\the\textfont2
  A\kern-.1667em\lower.5ex\hbox{M}\kern-.125emS}}

\title{
\vspace*{-35pt}
{\normalsize \hfill {\sf UTCCP-P-75}} \\
\vspace*{-6pt}
{\normalsize \hfill {\sf September \ 1999}} \\
Equation of state in finite-temperature QCD with improved Wilson quarks\thanks{presented by S.\ Ejiri
}
}

\author{CP-PACS Collaboration : 
      A.~Ali Khan\rlap,\address{Center for Computational Physics, University of Tsukuba, Tsukuba, Ibaraki 305-8577, Japan}
      S.~Aoki\rlap,\address{Institute of Physics, University of Tsukuba, Tsukuba, Ibaraki 305-8571, Japan}
      R.~Burkhalter\rlap,$^{\rm a,b}$
      S.~Ejiri\rlap,$^{\rm a}$
      M.~Fukugita\rlap,\address{Institute for Cosmic Ray Research,
      University of Tokyo, Tanashi, Tokyo 188-8502, Japan}
      S.~Hashimoto\rlap,\address{High Energy Accelerator Research Organization
      (KEK), Tsukuba, Ibaraki 305-0801, Japan}
      N.~Ishizuka\rlap,$^{\rm a,b}$
      Y.~Iwasaki\rlap,$^{\rm a,b}$
      K.~Kanaya\rlap,$^{\rm a,b}$
      T.~Kaneko\rlap,$^{\rm a}$
      Y.~Kuramashi\rlap,$^{\rm d}$
      T.~Manke\rlap,$^{\rm a}$
      K.~Nagai\rlap,$^{\rm a}$
      M.~Okamoto\rlap,$^{\rm b}$
      M.~Okawa\rlap,$^{\rm d}$
      H.P.~Shanahan\rlap,\address{DAMTP, University of Cambridge, Cambridge, CB3 9EW, England, UK}
      A.~Ukawa\rlap,$^{\rm a,b}$ and
      T.~Yoshi\'e$^{\rm a,b}$ }

\begin{document}

\begin{abstract}
We study finite-temperature phase transition and 
equation of state for two-flavor QCD at $N_t=4$
using an RG-improved gauge action 
and a meanfield-improved clover quark action. 
The pressure is computed using the integral method.
The O(4) scaling of chiral order parameter is also examined. 
\end{abstract}

\maketitle

\begin{figure}
\vspace*{-10mm}
\centerline{
\epsfxsize=8.2cm\epsfbox{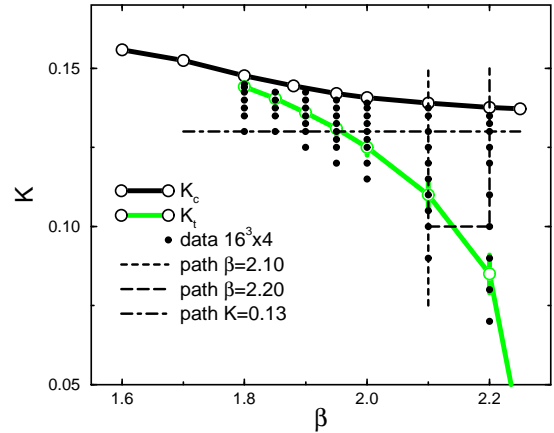}
}
\vspace*{-12mm}
\caption{
Phase diagram for $N_t=4$.
The solid line represents the critical line $K_c(\beta)$ of vanishing pion 
mass at $T=0$.  
The shaded line $K_t(\beta)$ is the location of finite-temperature 
transition. 
Dots represent the simulation points on $16^3\times4$ and $16^4$ lattices
carried out so far in the present work.
}
\label{fig:phase}
\vspace*{-2mm}
\end{figure}

\begin{figure}[t]
\vspace*{-5mm}
\hspace*{14mm}
\centerline{
\epsfxsize=8cm\epsfbox{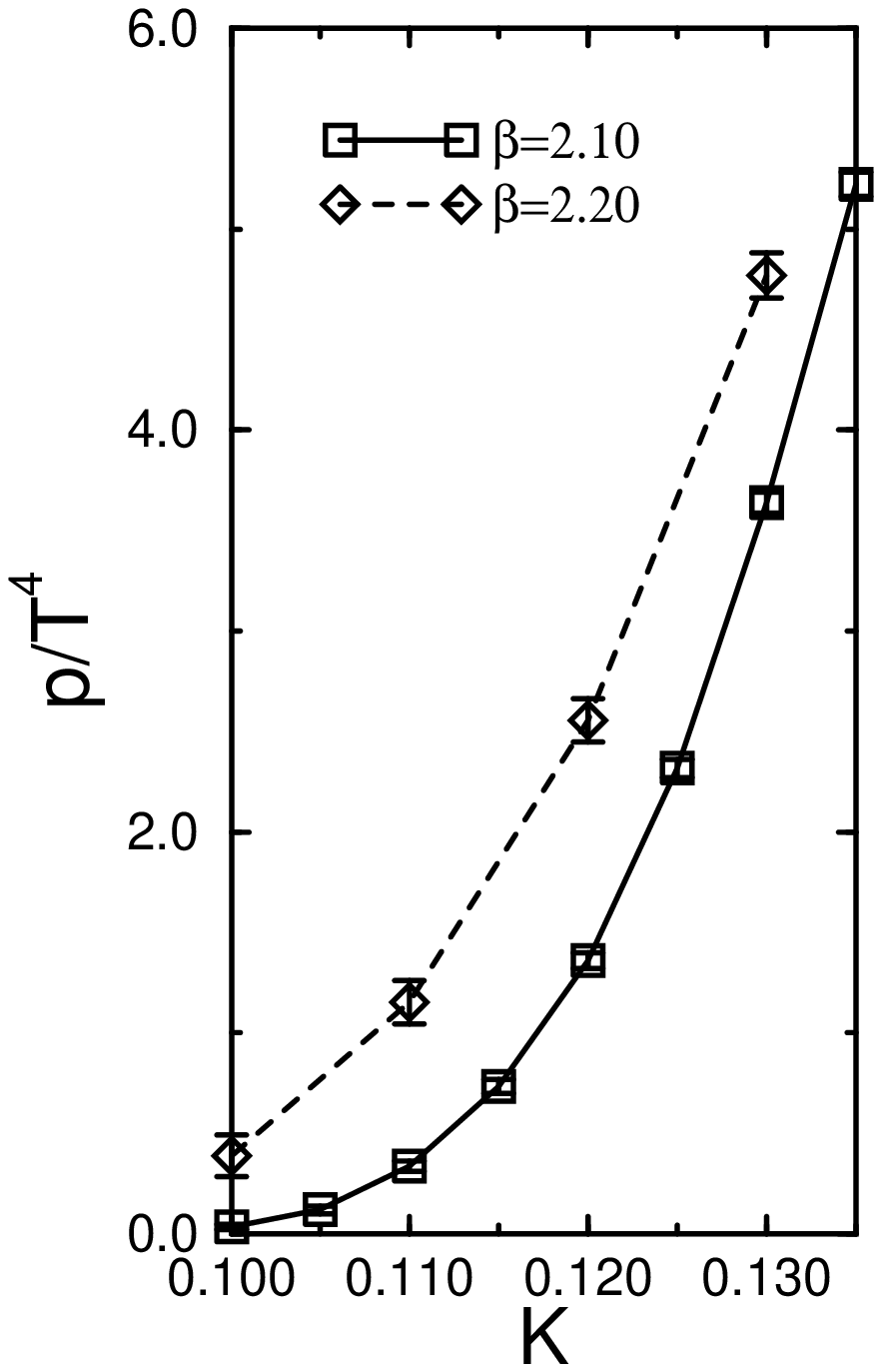}
\hspace*{-50mm}
\epsfxsize=8cm\epsfbox{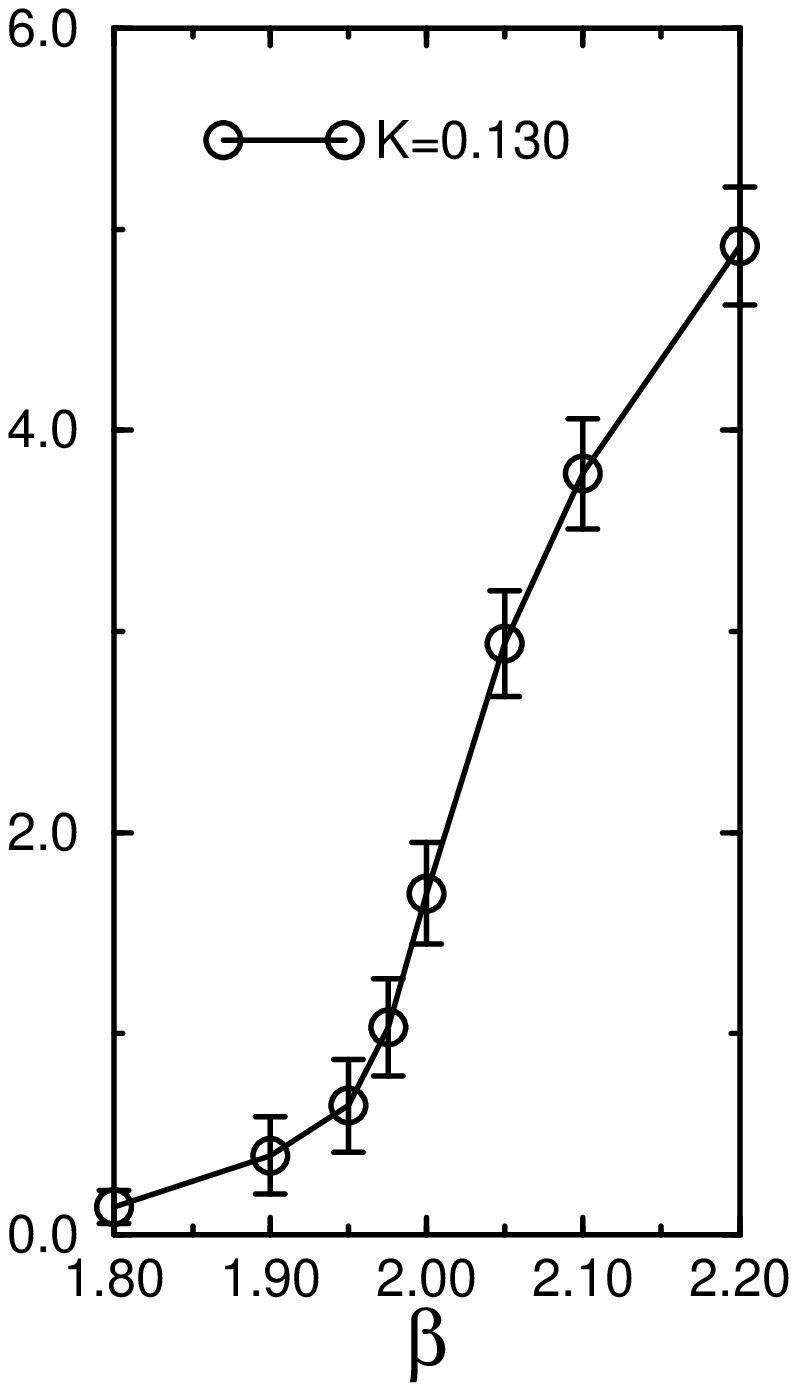}
}
\vspace*{-16mm}
\caption{Pressure on an $8^{3} \times 4$ lattice 
calculated from three paths in Fig.~\protect\ref{fig:phase}.}
\label{fig:prs84}
\vspace*{-2mm}
\end{figure}

\begin{figure}[t]
\vspace*{-5mm}
\centerline{
\epsfxsize=8.2cm\epsfbox{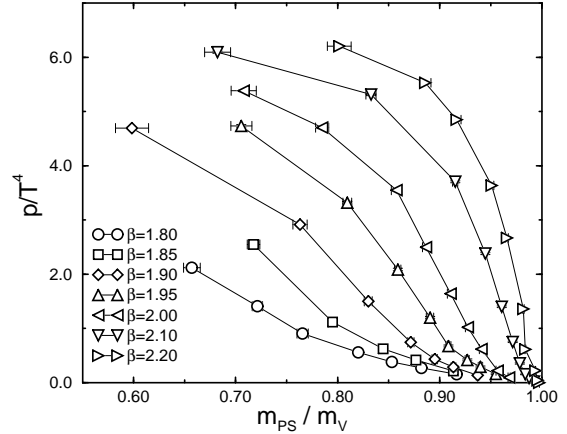}
}
\vspace*{-13mm}
\caption{
Pressure on a $16^3\times4$ lattice 
as a function of $\left. m_{\rm PS} / m_{\rm V}\right|_{T=0}$.
}
\label{fig:prs-pirho}
\vspace*{-2mm}
\end{figure}

\begin{figure}[t]
\vspace*{-5mm}
\centerline{
\epsfxsize=8.2cm\epsfbox{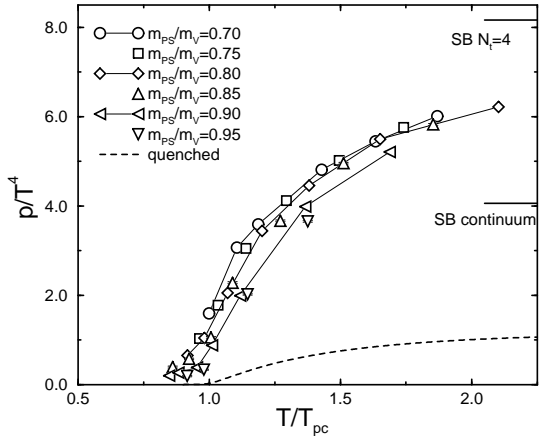}
}
\vspace*{-13mm}
\caption{
Pressure on a $16^3\times4$ lattice 
as a function of $T/T_{pc}$.
The dashed curve shows pressure
for pure gauge theory with the RG-improved action 
on a $16^3\times4$ lattice \protect\cite{okamoto}.
}
\label{fig:prs-ttc}
\vspace*{-2mm}
\end{figure}

\begin{figure}[t]
\vspace*{-5mm}
\centerline{
\epsfxsize=8.2cm\epsfbox{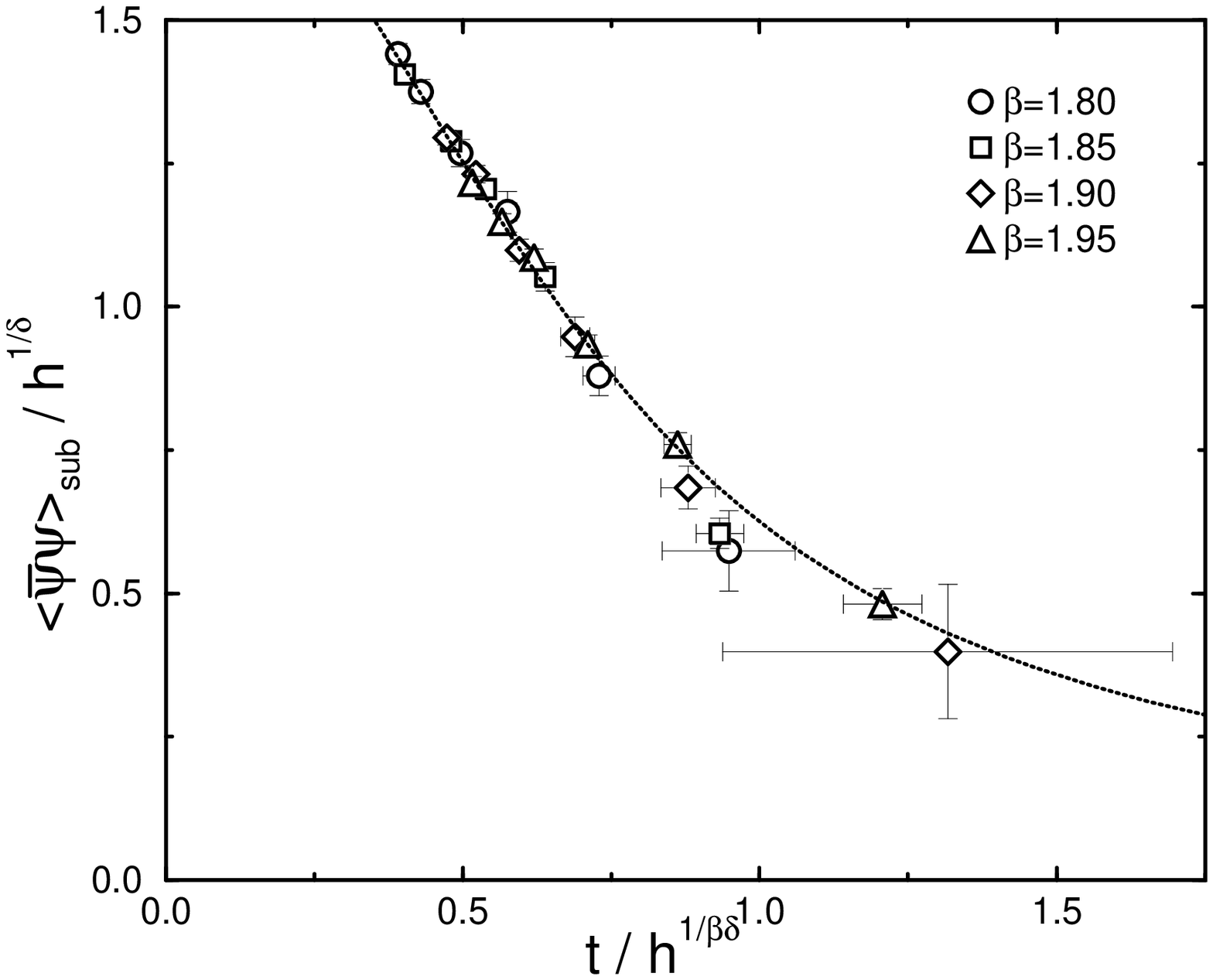}
}
\vspace*{-13mm}
\caption{
O(4) scaling relation.
}
\label{fig:o4}
\end{figure}

\section{Introduction}
\label{sec:intro}

The transition temperature and the equation of state (EOS) of QCD at 
finite temperature belong to the most basic information 
for understanding the early Universe and heavy ion collisions.
Full QCD studies of these quantities have been made mainly with 
the Kogut-Susskind quarks, particularly for the EOS \cite{karsch99}.
In this paper we present the first result of the EOS from 
Wilson-type quarks. 

We study two-flavor QCD on $N_t=4$ lattices.
In order to suppress lattice artifacts, which are known to be 
severe for the combination of the plaquette gauge and Wilson quark actions, 
we adopt a renormalization-group (RG) improved gauge action \cite{iwasaki} 
combined with a meanfield-improved clover quark action.
See Ref.~\cite{kaneko} for details of our action.

In Fig.~\ref{fig:phase} we show the phase diagram 
with our action at $N_{t}=4$. 
The line of finite-temperature transition 
is determined by the Polyakov loop and its susceptibility.
The parity-broken phase \cite{saoki} is not yet identified.
Dashed lines are used in a test discussed in Sec.~\ref{sec:eos}.

\section{Equation of state}
\label{sec:eos}

We compute the pressure $p$ by the integral method \cite{Eng90}, 
which is based on the formula, valid for large homogeneous systems,
that 
\begin{equation} 
\label{eq:integral}
\frac{p}{T^4} = - \frac{N_{t}^{3}}{N_{s}^{3}}
\int^{(\beta,K)} \!\!\!\!\!\!\!{\rm d} \xi \left\{ 
\left\langle \frac{\partial S}{\partial \xi} \right\rangle -
\left\langle \frac{\partial S}{\partial \xi} \right\rangle_{\!\! T=0} 
\right\} .
\end{equation}
The integration path in the parameter space $(\beta,K)$ should start from 
a point in the low temperature phase where the integrand 
approximately vanishes.
We evaluate the quark contributions to the derivatives 
$\frac{\partial S}{\partial \beta}$ and 
$\frac{\partial S}{\partial K}$ by the method of noisy source 
using U(1) noise vectors. 

The value for the pressure computed in (\ref{eq:integral}) 
should be independent of the choice of the integration path.
To check this point, 
we make a series of test runs on $8^3\times4$ and $8^{4}$ lattices
along three paths shown in Fig.~\ref{fig:phase}, 
generating 500 HMC trajectories at each point.
The results for $p/T^4$ obtained from these paths are plotted in 
Fig.~\ref{fig:prs84}.
We find that $p/T^4$ at $(\beta, K) = (2.1,0.13)$ and $(2.2,0.13)$ 
in the two figures are in good agreement, confirming the  
path independence of the integral. 

Encouraged by this result, we perform production runs on 
$16^{3} \times 4$ and $16^4$ lattices.
At each dots plotted in Fig.~\ref{fig:phase}, 
we generate 500--2000 trajectories on the $16^{3} \times 4$ lattice
and 200--300 trajectories on the $16^{4}$ lattice. 
Measurement of the derivatives is made at every trajectory.  
Hadron propagators are calculated at every fifth trajectory to compute   
pseudo scalar ($m_{\rm PS}$) and vector ($m_V$) meson masses.

As is seen from Fig.~\ref{fig:prs84}, paths along the $K$-direction 
give much smaller errors in $p/T^4$ 
than those from paths in the $\beta$-direction. 
Therefore, we carry out the integral in the $K$-direction. 
We then obtain pressure plotted in Fig.~\ref{fig:prs-pirho} 
as a function of the mass ratio 
$m_{\rm PS}/m_{\rm V}$ at zero temperature. 
Interpolating these data, we find $p/T^4$ for each value of 
$m_{\rm PS}/m_{\rm V}$.

Figure \ref{fig:prs-ttc} shows the pressure as a function of $T/T_{pc}$ 
at fixed $m_{\rm PS}/m_{\rm V}$.  
Here $T_{pc}$ is the pseudo-critical temperature at the same value of 
$m_{\rm PS}/m_{\rm V}$.  The temperature scale is set by $m_{\rm V}$ 
through $T/T_{pc}=m_{\rm V}(\beta_{pc})/m_{\rm V}(\beta)$ 
with $\beta_{pc}$ the pseudo-critical coupling.

We find that the pressure for fixed $T/T_{pc}$ depends only weakly 
on the quark mass even for relatively heavy quark in the 
range $m_{\rm PS}/m_{\rm V} = 0.7$--0.8.  
For heavier quark masses, the pressure decreases toward the pure gauge 
value (dashed line) \cite{okamoto} as expected.  

We also note that the magnitude of pressure is much larger than 
that for the pure gauge system for $N_t=4$, and that it overshoots 
the Stefan-Boltzman value in the continuum at high temperatures.  
These features are probably the result of large discretization errors
from the clover quark action here \cite{karsch99}.
Indeed the large Stefan-Boltzman value 
on an $N_t=4$ lattice shown at the top-right in Fig.~\ref{fig:prs-ttc} 
is dominated by the quark contribution.

\section{O(4) Scaling}
\label{sec:transition}

The chiral phase transition of two-flavor QCD is expected to 
belong to the universality class of O(4) spin system in three dimensions. 
In particular, 
identifying the magnetization, external magnetic field, and 
reduced temperature of the spin model 
with $M = \langle \bar{\Psi} \Psi \rangle$, $h = 2 m_{q} a$, 
and $t = \beta - \beta_{ct}$, 
where $\beta_{ct}$ is the chiral transition point, 
we expect a scaling relation
\begin{eqnarray}
M / h^{1/\delta} = f (t / h^{1/\beta \delta}) \label{scfn}
\end{eqnarray}
to hold with the O(4) scaling function $f(x)$ \cite{toussaint}
and the O(4) critical exponents $\beta$ and $\delta$ \cite{kaya}.
A previous study using the RG-improved gauge action and 
unimproved Wilson quark action \cite{o4scale} found this relation 
to be well satisfied for the quark mass and the chiral condensate 
defined by axial Ward identities \cite{bochicchio}. 

Figure \ref{fig:o4} shows the result of a similar analysis from 
the present work.  Data for 
$\langle \bar{\Psi} \Psi \rangle_{\rm sub}=2m_q a 
 (2K)^2 \sum_{x} \langle \pi (x) \pi(0) \rangle$
are fitted to the scaling relation, adjusting $\beta_{ct}$ and 
the scales for $t$ and $h$. 
The scaling ansatz works well, yielding $\beta_{ct} = 1.47(7) $
for the best fit of data for $2m_q a<0.9$ and $\beta\le 1.95$
with $\chi^{2}/{\rm df} = 1.1 $.

\vspace{2mm}

This work is in part supported by 
the Grants-in-Aid of Ministry of Education, Science and Culture 
(Nos.~09304029, 10640246, 10640248, 10740107, 
11640250, 11640294, 11740162). 
SE, KN and M.\ Okamoto are JSPS Research Fellows.
AAK and TM are supported by the Research for the Future 
Program of JSPS.

\end{document}